\documentstyle[prl,aps,preprint]{revtex}

\begin{document}
\renewcommand{\theequation}{\arabic{section}.\arabic{equation}}

\widetext
\draft
\title{
Disclination Asymmetry in Deformable Hexatic Membranes\\
and the Kosterlitz-Thouless Transitions
}
\author{Jeong-Man Park and T. C. Lubensky\\
Laboratory for Research on the Structure of Matter, \\
University of Pennsylvania, Philadelphia, PA 19104
}
\maketitle

\begin{abstract}
A disclination in a hexatic membrane favors the development of Gaussian
curvature localized near its core. The resulting global structure of
the membrane has mean curvature, which is disfavored by curvature energy.
Thus a membrane with an isolated disclination undergoes a buckling
transition from a flat to a buckled state as the ratio $\kappa/K_{A}$
of the bending rigidity $\kappa$ to the hexatic rigidity $K_{A}$ is
decreased.
In this paper we calculate the buckling transition and the energy of
both a positive and a negative disclination.
A negative disclination has a larger energy and a smaller critical value
of $\kappa/K_{A}$ at buckling than does a positive disclination.
We use our results to obtain a crude estimate of the Kosterlitz-Thouless
transition temperature in a membrane.
This estimate is higher than the transition temperature recently obtained
by the authors in a renormalization calculation.
\end{abstract}
\par
\section{Introduction}
\par
The hexatic phase \cite{hex-ph} is characterized by 
$6$-fold orientational but not
translational order.  Both three-dimensional hexatic phases with true 
long-range order \cite{3Dhex} and two-dimensional phases with power-law
order \cite{2Dhex} have been observed.  In flat two-dimensional films
(under tension), the transition to the isotropic phase occurs via a
Kosterlitz-Thouless (KT) disclination unbinding transition \cite{Kttran}.
Free membranes with zero surface tension can also exhibit a hexatic phase
and a KT transition to a fluid phase.  Both the hexatic phase and the
transition to the isotropic phase \cite{Nel-pel,Gui-kar}
are, however, more complicated than they are in a flat film
because of thermally induced 
shape fluctuations.   Recent renormalization group
calculations \cite{jp-lub1,jp-lub2} show that shape fluctuations shift the bare
hexatic rigidity $K_A$.  As a consequence, an increase in the amplituded
of shape fluctuations produced by decreasing the membrane bending rigidity
$\kappa$ will induce a transition from the hexatic phase to the isotropic
fluid phase.
\par
In flat membranes, there is a symmetry between positive ($5$-fold) and
negative ($7$-fold) disclinations, and they both have the same energy.  In
free deformable membranes, this symmetry is broken \cite{nel-pre}.  
If the ratio $\kappa / K_A$ of the bending rigidity $\kappa$ 
to the hexatic rigidity $K_{A}$ is sufficiently
small, a membrane with a single dislcination can lower its energy by
buckling, thereby creating a nonvanishing Gaussian curvature, which
screens the disclination charge.  The buckled states of positive and
negative disclinations have different height profiles, different
energies, and different critical values of $\kappa/K_A$ at which buckling 
occurs. In this paper, we will use a variational procedure to calculate
the energies of isolated positive and negative disclinations on free
membranes.  Our variational form for a positive disclination is essentially
exact.  Our form for negative disclinations is exact only
for $\kappa/K_A$ near the critical value of $\kappa/K_A$ for buckling.
Corrections near buckling are, however, very small, and we argue that our
form is a very good approximation until $\kappa/K_A$ becomes very small.
Our results are in agreement with recent calculations by Deem and Nelson
\cite{Deem-nel}. The latter authors, however, in addition to calculating
the energy of a negative disclination variationally also carry out
a numerical minimization of the full non-linear energy to obtain
a lower value for this energy at small values of $\kappa/K_{A}$.
In what follows, we will present our calculations of positive and negative
disclinations, respectively in Secs. II and III.  In Sec. IV, we will
discuss our results and an estimate of the $KT$ transition line produced
by them. 
\par
The continuum Hamiltonian for hexatic membranes was derived in
Ref. \cite{jp-lub1}.  If we parametrize membranes positions in terms of a
two-dimensional parameter ${\tilde u} = ( u^1, u^2)$ as ${\bf R}({\tilde
u})$ , then
\begin{equation}
{\cal H} = {\cal H}_{\kappa} + {\cal H}_C ,
\end{equation}
where
\begin{equation}
{\cal H}_{\kappa} = \frac{1}{2}\kappa\int d^{2}u\sqrt{g} H^{2},
\end{equation}
is the curvature energy and
\begin{equation}
{\cal H}_{C} =  \frac{1}{2}K_{A}\int d^{2}u\sqrt{g} 
              ({\cal N} - S)\frac{1}{-\nabla^{2}}
              ({\cal N} - S),
\end{equation}
is the Coulomb energy. 
Alternative form for the hexatic energy can be found in Ref. \cite{ref-david}.
In the above,
$g = \det ( g_{ab} )$ is the determinant of the metric tensor 
$g_{ab} = \partial_a {\bf R} \cdot \partial_b {\bf R} $ ,
${\cal N} = 2\pi\sum_{i}q_{i}\delta(\tilde{u}-\tilde{u}_{i})/\sqrt{g}$
is the disclination density with $q_{i} = \pm 1/6$, and
$H$ and $S$ are, respectively,
the mean and Gaussian curvatures of the membrane. 
We have ignored a scalar field contribution of ${\cal H}$, which gives
rise to Liouville measure factors \cite{Polya,Cai-lub}, 
which are irrelevant to the current discussion.
On a rigid flat membrane, this Hamiltonian reduces to the Coulomb gas form
of the XY-model.
In the absence of disclinations, 
hexatic order induces long-range Coulombic interaction between Gaussian
curvatures on the membrane \cite{Nel-pel}. The Coulomb energy ${\cal H}_C$
depends only on the difference ${\cal N} - S$.  Thus, the development of
Gaussian curvature on a free membrane that approximates the disclination
dentsiy ${\cal N}$ can reduce the Coulomb energy.  Gaussian curvature
usually leads to mean curvature, and the lowest energy state of a free
membrane with a disclination will be determined by the competition between
the Coulomb energy, ${\cal H}_C$, which prefers $S = {\cal N}$, 
and the curvarure energy,
${\cal H}_{\kappa}$, which prefers zero curvature.  If there is a single
disclination at the origin, one can expect that Gaussian curvature will be
localized near the origin.  Gaussian curvature localized in a small region
will give rise to a buckled state with mean curvature but zero Gaussian
curvature away from the origin.    
\setcounter{equation}{0}
\section{Positive Disclinations}
A minimum strength positive disclination has ``charge" $q = 1/6$.  To
reduce the Coulomb energy associated with this charge, the membrane can
distort into the shape of a spherical section, with
nonvanishing Gaussian curvature, localized to the core of the
disclination. Outside the core region, the membrane will seek a shape with
zero Gaussian curvature.  A cone with slope $m$ has zero Gaussian
curvature and can be connected smoothly to a spherical section 
(Fig.~\ref{cone}). Thus, in the
Monge gauge, we parametrize the membrane shape outside the core as
${\tilde u} = (r, \phi)$ and ${\bf R}({\tilde u}) = (r\cos\phi, r\sin\phi
, h ( {\tilde u} ))$ with $h( {\tilde u} ) = m r$.  
Thus, outside the core, the components of metric tensor and its determinant 
are 
\begin{equation}
\begin{array}{cc}
g_{rr} = 1 + m^2 , & \qquad\qquad g_{\theta\theta} = r^2, \\
g_{r\theta} = g_{\theta r} = 0 , & \qquad\qquad g = r^2 (1 + m^2). 
\end{array} 
\label{posg}
\end{equation}
The mean curvature $H$ and the Gaussian curvature $S$ are 
\begin{equation}
H = \frac{m}{\sqrt{1+m^{2}}}\frac{1}{r}, \;\;\; S = 0 \;.
\end{equation}
Thus the bending energy of the cone with radius $R$ and the core size $a$
becomes
\begin{eqnarray}
{\cal H}_{\kappa} & = & \frac{1}{2}\kappa\int d^{2}u\sqrt{g}H^{2} \nonumber  \\
   & = & \frac{1}{2}\kappa\int_{a}^{R}\frac{2\pi rdr}{r^{2}}\sqrt{1+m^{2}}
         \left( \frac{m}{\sqrt{1+m^{2}}} \right)^{2}  \nonumber  \\
   & = & \pi\kappa\frac{m^{2}}{\sqrt{1+m^{2}}}\ln\frac{R}{a}.
\end{eqnarray}
The Gaussian curvature vanishes outside the core region. 
In the limit of the infinitesimal size of the core region, $S$ can
be described by the point curvature charge $s_{+}$:
\begin{equation}
S(\tilde{u}) = 2\pi s_{+}\delta(\tilde{u}-\tilde{u}_{+})/\sqrt{g}.
\label{Su}
\end{equation}
Since there is no Gaussian curvature in the cone, we can choose any curve
in the cone to calculate $s_{+}$ using the Gauss-Bonnet theorem:
\begin{equation}
\int_{M} S d\sigma +\int_{C}k_{g} dl = 2\pi,
\end{equation}
where $k_{g}$ is the geodesic curvature of the boundary curve $C$
of the surface $M$ \cite{dubro2}.
We use the boundary curve of the cone:
\begin{equation}
C = (R\cos\phi, R\sin\phi, mR).
\end{equation}
The geodesic curvature of the boundary curve of a cone of slope $m$ 
is $(R\sqrt{1+m^{2}})^{-1}$, and
the Gauss-Bonnet theorem becomes
\begin{equation}
\int_{M} S d\sigma + \int_{0}^{2\pi}\frac{d\phi}{\sqrt{1+m^{2}}} 
= 2\pi.
\end{equation}
Using Eq.\ (\ref{Su}), we obtain
\begin{equation}
s_{+} = 1- \frac{1}{\sqrt{1+m^{2}}}.
\end{equation}
Hence, the Coulomb energy for a positive disclination is 
\begin{eqnarray}
{\cal H}_{\rm C} & = & \frac{1}{2}K_{A}\int d^{2}u\sqrt{g}
                ({\cal N}-S)\frac{1}{-\nabla^{2}}({\cal N}-S) \nonumber  \\
   & = & \pi K_{A}\left(\frac{1}{6}-s_{+}\right)^{2} 2\pi G_{c}(0),
\end{eqnarray}
where $s_{+}=1-1/\sqrt{1+m^{2}}$ and $G_{c}(\tilde{u})$ is the 
Green's function for the Laplacian,
\begin{equation}
\nabla^{2} = \frac{1}{\sqrt{g}}\partial_{\alpha}g^{\alpha\beta}\sqrt{g}
\partial_{\beta} \;.
\end{equation}
On a cone,
\begin{equation}
\nabla^{2}G_{c}(\tilde{u}-\tilde{u}') = 
- \delta(\tilde{u}-\tilde{u}')/\sqrt{g}.
\label{pos-gf}
\end{equation}
To determine $G_c$, we assume that it has the form $-A\ln(r/r_0)$ where
$r_0$ is a length.  Then
\begin{eqnarray}
\int d^2 u \sqrt{g} \nabla^2 G_c ( {\tilde u} ) & = & - \int d^2 u
\sqrt{g} \delta( {\tilde u})/\sqrt{g} = -1 \nonumber \\
& = & \int ds_a g^{ab} \sqrt{g} \partial_b G_c = - \int ds_r g^{rr} A/r
\nonumber \\
& = & - \int ds_r (g_{\theta \theta} / \sqrt{g} ) (A/r ) ,
\label{Gcu}
\end{eqnarray}
where $ds_a = \delta_{ar} rd\theta$ is the ``surface" element of a circle
enclosing the orgin.  Then from Eq.\ (\ref{posg}),
$g_{\theta\theta}/\sqrt{g} = r/\sqrt{1 + m^2}$, $A = -\sqrt{1 + m^2}/(2
\pi)$, and
\begin{equation}
G_{c}(\tilde{u}) = -\frac{\sqrt{1+m^{2}}}{2\pi} \left( \ln
\frac{r}{a} - \ln\frac{R}{a} \right) ,
\end{equation}
where we chose $r_0$ to be equal to the disclination core radius $a$ and
we added the constant term $-\ln(R/a)$, where $R$ is the radius of the cone,
to produce the required divergence of $G_{c}(\tilde{u})$ at small $r$.
The Coulombself-energy is given in terms of 
$2\pi G_{c}(0) = \sqrt{1+m^{2}}\ln(R/a)$
where $R$ is the radius of the membrane and $a$ is the core size.
\par
\begin{figure}
\centerline{\psfig{figure=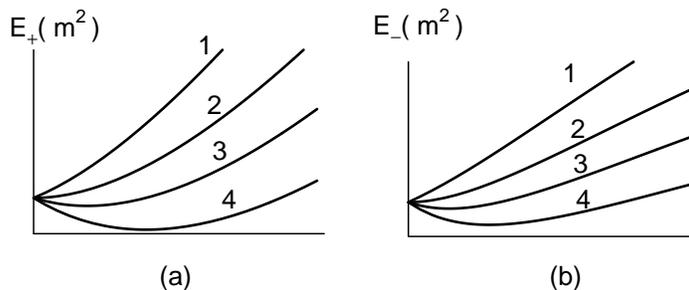}}
\caption{(a) Energy of positive disclinations as a function of $m^{2}$
for different values of $\rho= \kappa/K_{A}$:
(1) $\rho/\rho_{c}^{+} = 4/3$, (2) $\rho/\rho_{c}^{+} = 1$, 
(3) $\rho/\rho_{c}^{+} = 3/4$, and (4) $\rho/\rho_{c}^{+} = 1/2$, where
$\rho_{c}^{+}=11/72$. 
(b) Energy of negative disclinations as a function of $m^{2}$
for different values of $\rho= \kappa/K_{A}$:
(1) $\rho/\rho_{c}^{-} = 4/3$, (2) $\rho/\rho_{c}^{-} = 1$, 
(3) $\rho/\rho_{c}^{-} = 3/4$, and (4) $\rho/\rho_{c}^{-} = 1/2$, where
$\rho_{c}^{-}=13/216$. 
}
\label{ener}
\end{figure}

The energy of a positive disclination with $q=1/6$ 
on the cone with the slope $m$ becomes
\begin{eqnarray}
E_{+}(m) & = & \left[ \pi\kappa\frac{m^{2}}{\sqrt{1+m^{2}}} 
+ \pi K_{A} \left(
\frac{1}{6} - 1 + \frac{1}{\sqrt{1+m^{2}}} \right)^{2} \sqrt{1+m^{2}}
\right] \ln\frac{R}{a}    \nonumber  \\
   & \simeq & \pi K_{A} \left[ \frac{1}{36} + \left( \frac{\kappa}{K_{A}}
- \frac{11}{72} \right) m^{2} + \left( \frac{83}{288} - \frac{1}{2}
\frac{\kappa}{K_{A}} \right) m^{4} \right] \ln \frac{R}{a}.
\label{energy+}
\end{eqnarray}
This energy is shown in Fig.~\ref{ener}(a) for various values of
$\kappa/K_{A}$.
For $\kappa/K_{A} > 11/72$, $E_{+}(m)$ has a minimum at
$m^{2}=0$ and the membrane remains flat.
For $\kappa/K_{A} < 11/72$, however, $E_{+}(m)$ has a minimum at
$m^{2} = (11/36-2\kappa/K_{A})/(25/36+\kappa/K_{A})$ and the membrane 
buckles out to form a cone with the slope 
\begin{equation}
m = 
\pm\sqrt{\frac{\frac{11}{36}-\frac{2\kappa}{K_{A}}}
{\frac{25}{36}+\frac{\kappa}{K_{A}}}}.
\end{equation}
Thus the buckling transition occurs at $\kappa/K_{A} = 11/72$ for positive
disclination with $q=1/6$.
This result for $m$ with Eq. \ref{energy+} can be found in Ref. 
\cite{ref-guitter} and \cite{ref-seung}.
The energy of a positive disclination on a membrane of radius $R$ with
the short-distance cutoff $a$ is
\begin{equation}
E_{+} = \left\{ \begin{array}{ll}
 \frac{5}{3}\pi K_{A}\left( \sqrt{\left(1-\frac{\kappa}{K_{A}}\right)
\left(1+\frac{36\kappa}{25K_{A}}\right)} - 1 \right)\ln\frac{R}{a}, &
             \frac{\kappa}{K_{A}} < \frac{11}{72}   \\
 \frac{1}{36}\pi K_{A}\ln\frac{R}{a}, & \frac{\kappa}{K_{A}} > \frac{11}{72}.
               \end{array}
        \right .
\end{equation}
When $K_A \rightarrow \infty$, $m \rightarrow \pm \sqrt{11/25}$ and $s_+
\rightarrow 1/6$.  Thus, in this limit, the disclination charge is totally
screened by the Gaussian curvature, there is no Coulomb energy, and the
disclination energy $E_+ ( K_A = \infty ) = (11/30) \pi \kappa \ln
(R/a)$ comes entirely from curvature of the cone.
This energy of a positive disclination is identical to the result obtained
by Guitter and Kardar \cite{Gui-kar} using the conformal gauge.
\par
Our simple height profile for a positive disclination does not break azimuthal
symmetry, and there is no particular reason for this symmetry to be broken.
Thus, we believe that $h(\tilde{u})=mr$ provides a complete discription of
the buckled state and our description of the positive disclinations is exact.
In particular, no symmetry breaking terms such as $m_{1}r\cos 2\phi$ or
$m_{2}r\cos 4\phi$ are needed in the expansion of $h(\tilde{u})$.
Order parameters such as $m_{1}$ and $m_{2}$ are certainly not forced by
the development of non-zero $m$ because symmetry does not permit terms
linear in $m_{p}$ of the form $m^{k}m_{p}$ to appear in the expansion
of $E_{+}(m)$.
\par
A KT melting temperatue $T_{+}$ for positive disclinations can be introduced
in the usual way by setting the free energy of a single disclination equal
to zero; $E_+ - T_+ {\cal S}=0$, where ${\cal S} = \ln (R/a)^2$ is the
entropy.  Thus $T_+ = E_+/2 \ln (R/a)$.  This produces the phase diagram
obtained in Ref.\ \cite{Gui-kar} and shown as the solid curve (a) 
in Fig.~\ref{diag1}.  
The thin line indicates the buckling transitions at $\kappa/K_{A}=11/72$, 
and the solid curve the
disclinations unbinding transtion obtained from $T_+$.
\setcounter{equation}{0}
\section{Negative Disclinations}
We can similarly calculate the energy of negative disclinations with 
$q=-1/6$. Since the corresponding core region should have a
negative curvature charge to cancel the topological charge, we expect the
core region has a saddle shape.
The simplest saddle shape (Fig.~\ref{saddle}) is 
\begin{equation}
h(\tilde{u}) = mr\cos 2\phi \;.
\end{equation}
We will take this as a variational function and seek the minimum
energy solution for a negative disclination with respect to variations
in the parameter $m$.
We thus obtain an upper bound to the energy of a negative disclination.
Inclusion of additional terms in $h(\tilde{u})$ proportional to
$\cos 2 n \phi$ for $n$ an integer will lead to lower energies.
Indeed, recent numerical calculations by Deem and Nelson \cite{Deem-nel} 
yield a lower energy than we obtain when $K_{A}/\kappa \gg 1$.
We will argue, however,
that this simple variational form is essentially exact near
the buckling transition.
\par
The components of the
metric tensor and its determinant 
associated with $h(\tilde{u}) = mr\cos 2\phi$ are
\begin{equation}
\begin{array}{cc}
g_{rr} = 1 + m^2 \cos^2 2 \phi , &
               g_{\theta\theta} = r^2 ( 1 + r m^2 \sin^2 2 \phi ), \\
g_{r\theta} = g_{\theta r} = - 2 r m^2 \cos 2 \phi \sin 2 \phi , &
\;\;\; g = r^2 ( 1 + m^2 ( 1 + 3 \sin^2 2 \phi )) .
                  \end{array} 
\end{equation}
The mean curvature $H$ and the Gaussian curvature $S$ are 
\begin{equation} 
H = -\frac{3m}{r}\cos 2\phi 
\frac{1+m^{2}\cos^{2}2\phi}{(1+m^{2}(1+3\sin^{2}2\phi))^{3/2}}, \;\;\; S = 0.
\end{equation}
The bending energy of the saddle with slope $m$ is 
\begin{eqnarray}
{\cal H}_{\kappa} & = & \frac{1}{2}\kappa\int d^{2}u\sqrt{g}H^{2} \nonumber  \\
   & = & \frac{1}{2}\kappa\int_{a}^{R}\frac{rdr}{r^{2}}
    \int_{0}^{2\pi}d\phi\frac{9m^{2}\cos^{2}2\phi(1+m^{2}\cos^{2}2\phi)^{2}}
         {(1+m^{2}(1+3\sin^{2}2\phi))^{5/2}}  \nonumber  \\
   & = & \frac{9}{2}\kappa m^{2} \left[ \int_{0}^{2\pi}d\phi
         \frac{\cos^{2}2\phi(1+m^{2}\cos^{2}2\phi)^{2}}
         {(1+m^{2}(1+3\sin^{2}2\phi))^{5/2}} \right] \ln\frac{R}{a}.
\label{bend-neg}
\end{eqnarray}
Again, in the limit of the infinitesimal size of the core region,
$S$ can be described by the point curvature charge $s_{-}$:
\begin{equation}
S(\tilde{u}) = 2\pi s_{-}\delta(\tilde{u}-\tilde{u}_{-})/\sqrt{g}.
\end{equation}
The integrated geodesic curvature along the boundary 
$C = (R\cos\phi, R\sin\phi, mR\cos 2\phi)$ is 
\begin{equation}
\int_{C} k_{g}dl = \int_{0}^{2\pi}\frac{(1+4m^{2})d\phi}
{(1+m^{2}\cos^{2}2\phi)^{1/2}(1+4m^{2}\sin^{2}2\phi)^{3/2}} .
\end{equation}
Thus the Gauss-Bonnet theorem gives
\begin{equation}
s_{-} = 1 - \frac{1}{2\pi}\int_{0}^{2\pi}\frac{(1+4m^{2})d\phi}
{(1+m^{2}\cos^{2}2\phi)^{1/2}(1+4m^{2}\sin^{2}2\phi)^{3/2}},
\label{s-neg}
\end{equation}
and the Coulomb self-energy of negative disclination on the saddle becomes
\begin{eqnarray}
{\cal H}_{\rm C} & = & \frac{1}{2}K_{A}\int d^{2}u\sqrt{g}
                ({\cal N}-S)\frac{1}{-\nabla^{2}}({\cal N}-S) \nonumber  \\
   & = & \pi K_{A}\left(\frac{1}{6}+s_{-}\right)^{2} 2\pi G_{s}(0),
\label{H_C}
\end{eqnarray}
where $s_{-}$ is given by Eq. (\ref{s-neg}) and $G_{s}(\tilde{u})$
is the Green's function for the Laplacian $\nabla^{2}$ on the saddle.
We can determine $G_s ( {\tilde u})$ following exactly the same procedure
we used to determine $G_c ( {\tilde u})$ for a positive disclination.  We
find
\begin{equation}
G_s ( {\tilde u} ) = - A [ \ln (r/a) - \ln(R/a) ]
\label{G_s}
\end{equation}
where
\begin{eqnarray}
A & = &\left(\int d \phi {1 + 4 m^2 \sin^2 2 \phi \over [1 + m^2 (1 + 3
\sin^2 2 \phi )]^{1/2}}\right)^{-1} \nonumber \\
& \simeq & {1 \over 2 \pi} \left[1 - {3 \over 4} m^2 + {67 \over 64}
m^4 + \cdots \right] .
\label{AG_s}
\end{eqnarray}
Equations (\ref{bend-neg}), (\ref{H_C}), (\ref{G_s}), and (\ref{AG_s})
completely determine the energy of a negative discliantion as a function
of $m$ within the approximation $h({\tilde u}) = m r \cos 2 \phi$.  We can
locate the buckling instability and determine $m$ just above it by
expanding $E_{-}(m)$ in powers of $m$ up to order $m^4$.  The result is
\begin{equation}
E_{-}(m) = \pi K_A \left[ \frac{1}{36} + 
{9\over 2}\left( {\kappa\over K_A} - \frac{13}{216}
\right)m^{2} + \left( \frac{2743}{2306} - \frac{207}{16}
{\kappa \over K_A} \right)
m^{4} \right] \ln\frac{R}{a} .
\end{equation}
This energy is shown in Fig.~\ref{ener}(b) for various values of
$\kappa/K_{A}$.
For $\kappa/K_{A} > 13/216$, $E_{-}(m)$ has a minimum at $m^{2}=0$;
for $\kappa/K_{A} < 13/216$, $E_{-}(m)$ has a minimum at $m^{2}=
((13/216)-\kappa/K_{A})/((2743/10368)-(23/4)\cdot\kappa/K_{A})$.
The buckling transition occurs at 
\begin{equation}
{\kappa \over K_A} = {13 \over 216} ,
\end{equation}
and the slope for $\kappa/K_{A} < 13/216$ is
\begin{equation}
m = \pm\sqrt{\frac{(13/216)-\kappa/K_{A}}
{(2743/10368)-(23/4)\cdot\kappa/K_{A}}}.
\label{m-neg}
\end{equation}

The energy of negative disclination around $\kappa/K_{A} = 13/216$ 
in a membrane of radius $R$ with the 
short-distance cutoff $a$ is
\begin{equation}
E_{-} = \left\{ \begin{array}{ll}
 \frac{\pi}{36} K_{A} \left( 1-\frac{9((13/48)-(9/2)(\kappa/K_{A}))^{2}}
 {((2743/2304)-(207/16)(\kappa/K_{A}))}\right)\ln\frac{R}{a}, &
             \frac{\kappa}{K_{A}} < 13/216   \\
 \frac{\pi}{36} K_{A}\ln\frac{R}{a}, & \frac{\kappa}{K_{A}} > 13/216.
               \end{array}
        \right.
\label{e-neg}
\end{equation}
As in the case of positive disclinations, the Gaussian curvature will
adjust to exactly cancel the topological charge when $K_A = \infty$,
leaving only curvature energy.  Setting $s_-(m)$ in Eq.\ (\ref{s-neg})
equal to $-1/6$, we obtain
\begin{equation}
m^{2}(K_A = \infty) = 0.350417
\end{equation}
and
\begin{equation}
E_{-}(K_A = \infty ) = 2.75883\kappa \ln\frac{R}{a}.
\end{equation}
$E_-(m)$ can be minimized numerically for $0 < \kappa / K_A < 13/216$.
The results are displayed as a transition temperature in Fig.~\ref{diag1}
(See below).
\par
As discussed in the introduction, the height profile of a negative
disclination breaks azimuthal symmetry, and we expect $h ({\tilde u})$ to
have a Fourier series expansion of the form $h( {\tilde u}) = r\sum_n m_n
\cos 2 n \theta$.  Our approximation keeps only the first term in this
series.  Near the transition, higher order terms can be calculated by
expanding $E_-$ in a powers series in all of the $m_n$'s.  We have already
calculated the contribution from the dominant term $m_1 \equiv m$.  One
might expect that the next most important term would be $m_2$.  This
parameter is not, however, forced to develop a nonzero value when $m$ is 
nonzero because $E_-$ is invariant under $h \rightarrow -h$, i.e., under
$m_n \rightarrow - m_n$ for every $n$.  Thus, there is a contribution to
$E_{-}$ of the form $a_2 m_2^2$ but no term
proportional to $m^2 m_2$, which would force a nonzero $m_2$.  
Thus $m_2$ will remain zero until the coefficient $a_2$ changes sign.  The
absence of an $m^2 m_2$ term means that our expressions for 
$E_-$ [Eq.\ (\ref{e-neg})] and $m$ [Eq.\ (\ref{m-neg})] are exact to order 
$[(\kappa/K_A) - (13/216)]^2$ because there is no correction to the $m^4$
term arising from couplings to $m_2$.  The third order term $m_3$ is
proportional to $m^3$ because inversion symmetry permit a term
proportional to $m^3 m_3$.  More generally, there are couplings of the
fomr $m^{2 p + 1} m_{2 p +1}$ for $p = 1 , 2 , ...\;$.  Thus the height
profile can be expanded as $h(r, \theta) = r \sum_{p = 0} m_{2 p + 1} \cos
[2 (2 p + 1 ) \theta]$.  Our results for $E_-$ and $m$ agree with those
obtained analytically and numerically by Deem and Nelson \cite{Deem-nel}.  
Their numerical
result for $\kappa /K_A = 0$ is lower than our indicating that the order
parameters $m_{2 p +1}$ for $p \geq 1$ are important in this limit.
\par
A KT transition temperature for negative discliantion can be introduced
just as for positive disclinations: $T_- = E_-/2 \ln (R/a)$.  
This curve is shown as the solid curve (b) in Fig.~\ref{diag1}.
\setcounter{equation}{0}
\section{Discussion}
\begin{figure}
\centerline{\psfig{figure=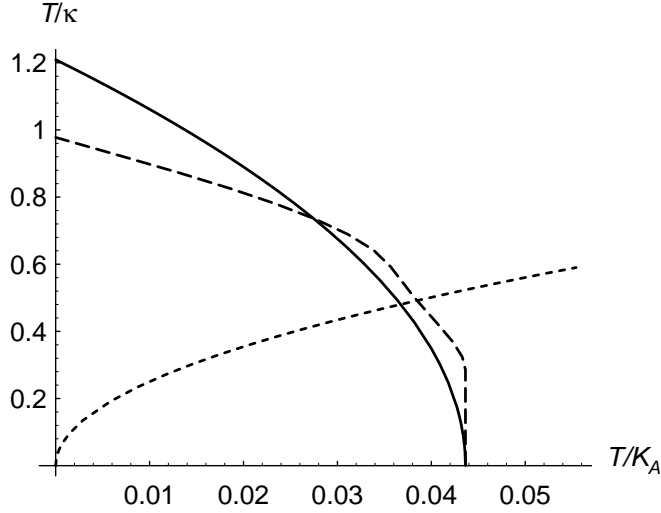}}
\caption{
Phase diagram in $(T/\kappa, T/K_{A})$ plane.
The full line is the KT transition line obtained from the RG calculation of
Refs. \protect{\cite{jp-lub1}} and \protect{\cite{jp-lub2}}.
The dashed line is the estimate of the KT transition line from the average
$(T_{+}+T_{-})/2$ of the positive and negative melting temperatures
$T_{+}$ and $T_{-}$. The approximations used in 
Refs. \protect{\cite{jp-lub1}} and \protect{\cite{jp-lub2}} apply below
the dotted line. The RG transition line (full line) obtained in
Refs. \protect{\cite{jp-lub1}} and \protect{\cite{jp-lub2}} lies below
the simple estimate (dashed line) in the region (below the dotted line)
where the RG calculation applies.
This is expected since the RG calculation includes contributions to the
free energy arising from thermal membrane fluctuations which the simple
estimate $(T_{+}+T_{-})/2$ does not include.
}
\label{diag2}
\end{figure}

Clearly both positive and negative disclinations will be thermally
excited, and neither $T_{+}$ nor $T_{-}$ is a good estimate of the actual
melting temperature, $T_{M}$.
A better estimate is that $T_{M}$ is simply the average $(T_{+}+T_{-})/2$.
This yields the dashed curve in Fig.~\ref{diag2}.
This estimate describes qualitatively features that are in agreement
with simple physical reasoning: for large $\kappa$, there should be
a disclination-mediated melting to the disordered phase as temperature is
increased, and at fixed $K_{A}$, there should be a transition to the
crumpled phase as $\kappa$ is decreased.
In Refs. \cite{jp-lub1} and \cite{jp-lub2}, we calculated the melting 
temperature using the
renormalization group (RG) recursion relations for the KT transition on a
fluctuating membrane subject to the constraint of charge neutrality.
The result shown as the solid curve in Fig.~\ref{diag2} is below the 
estimate $(T_{+}+T_{-})/2$ in the entire region
below the dotted curve where we believe that our RG calculations are valid.
This is entirely reasonable. The estimate of $T_{M}$ obtained by equating
the energy of disclinations to temperature times their entropy completely
ignores the entropy associated with height fluctuations, which should lead 
to a depression of $T_{M}$.
Our RG calculations include height fluctuations, whose major effect is to
decrease the effective long-wavelength dielectric constant.
\par
We are grateful to David Nelson for bringing to our attention the asymmetry
in positive and negative disclination energies and to Michael Deem and 
David Nelson for sending us their independent work on this subject prior
to publication. This work was supported in 
part by the National Science Foundation under grant No. DMR94-23114 and
in part by the Penn Laboratory for Research on the Structure of Matter
under NSF grant No. DMR91-20668.
J.M.P. would like to give a special thanks to Prof. M.L. Klein 
for his kindness.

\input{psfig}
\newpage
\begin{figure}
\caption{Membrane buckled into a cone with $h(\tilde{u}) = mr$ in the 
Monge gauge where $m$ is the slope.
}
\label{cone}
\end{figure}
\begin{figure}
\centerline{\psfig{figure=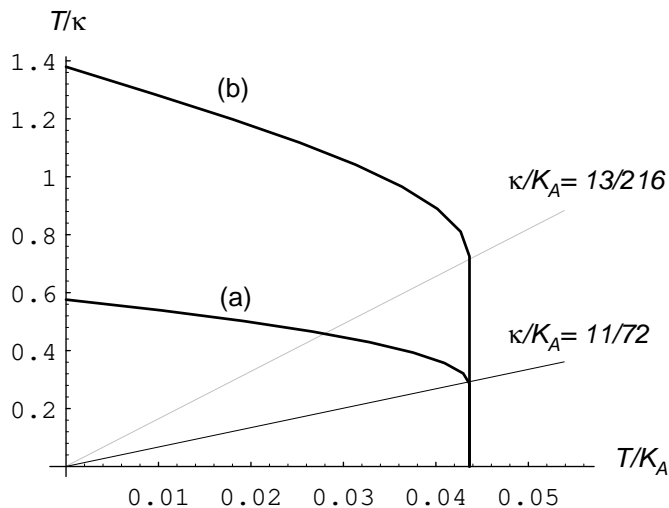}}
\caption{
Estimated phase diagrams in the $(T/\kappa, T/K_{A})$ plane showing the
Kosterlitz-Thouless transition line obtained by balancing energy and entropy 
of (a) a single positive disclination ($T_{+}=E_{+}/2\ln (R/a)$) 
and (b) a single negative disclination ($T_{-}=E_{-}/2\ln(R/a)$).
(a) is identical to the estimate obtained in Ref. \protect{\cite{Gui-kar}}.
The straight line through the origin in both cases is the buckling transition
line.
The energy of a negative disclination is generally higher than that of
a positive disclination, and $T_{-} > T_{+}$.
$T_{+}(K_{A}=\infty) = (11/60)\pi\kappa \simeq 0.575959\kappa$ and
$T_{-}(K_{A}=\infty) \simeq 1.37941\kappa$.
}
\label{diag1}
\end{figure}
\begin{figure}
\caption{Membrane buckled into a saddle with $h(\tilde{u}) = mr\cos 2\phi$
in the Monge gauge where $m$ is the slope.
}
\label{saddle}
\end{figure}
\end{document}